\documentclass[twoside,fleqn]{ActaStyle}
\usepackage{times,cite}
\usepackage{graphicx,epsfig}
\def\authorheading{A. Gendiar et al.}
\def\shorttitle{Estimation of the magnetic critical exponent}
\begin{document}

\pagerange{1}{8}

\title{ESTIMATION OF THE MAGNETIC CRITICAL EXPONENT BY
TENSOR PRODUCT VARIATIONAL APPROACH}

\author{
A. Gendiar,\email{gendiar@savba.sk}$^{,a}$ T. Nishino,$^b$ R. Derian$^c$
}
{
$^a$Institute of Electrical Engineering, Slovak Academy of Sciences, D\'{u}bravsk\'{a}
cesta 9\\ SK-841 04 Bratislava, Slovakia\\
$^b$Department of Physics, Faculty of Science, Kobe University, Kobe 657-8501, Japan\\
$^c$Institute of Physics, Slovak Academy of Sciences, D\'{u}bravsk\'{a}
cesta 9\\ SK-845 11 Bratislava, Slovakia\\}

\abstract{%
A variational problem for three-dimensional (3D) classical lattice models is considered
with trial state given by two-dimensional (2D) uniform product of local variational
weights. This approach, the tensor product variational approach (TPVA), has been applied
to 3D classical models (the Ising and the Potts models). We consider a way of estimating
the magnetic critical exponent $\beta$ for the simple 3D Ising model assuming a functional
form of the spontaneous magnetization in the off critical region, where the TPVA provides
reliable data.}

\pacs{05.70.Fh, 64.70.Rh, 02.60.Dc, 02.70.-c}

\section{Introduction}

Extension of Density Matrix Renormalization Group
(DMRG)~\cite{White1,White2,DMRG,Nishino} to higher dimensions is still of
main interest because a very slow decay in the density matrix eigenvalue
spectrum prevents the creation of efficient renormalization group 
transformation~\cite{Peschel}.

The numerical efficiency of the DMRG for the low-dimensional
systems can be explained from the variational background~\cite{Ost,Rom,Sierra}.
In DMRG, the variational state is constructed by the product of orthogonal
matrices. This type of the variational state can be defined in any dimension.
In the three-dimensional case, Nishino {\it et al.} proposed the tensor product
variational approach (TPVA)~\cite{TPVA1}. It is possible to regard the TPVA as
an extension of the DMRG to 3D classical (or 2D quantum) systems.

The TPVA has been mostly applied to 3D classical spin models such as
the Ising and the Potts models~\cite{TPVA1,TPVA2,TPVA3,TPVA4},
where we considered the spatially uniform spin systems. We have
also applied it to the 2D quantum Heisenberg model~\cite{Nishio}.
The TPVA is of use for the spin models which exhibit
ordered magnetic structures such as in the
3D Axial-Next-Nearest-Neighbor Ising model, where the trial
state is constructed from the position dependent tensors~\cite{ANNNI}. 
Such a position dependence is also considered by Verstraete {\it et al.}
for 2D quantum systems~\cite{Ver1,Ver2,Ver3}.

For the models which exhibit the second-order phase transition, the TPVA
tends to overestimate the critical temperature. For example, the error in
$T_{\rm c}$ is about 1\% or less for the 3D Ising model compared with the
Monte Carlo (MC) result~\cite{MC}. This is partially because the
correlation length diverges at the criticality and the TPVA cannot
treat this divergence properly due to a finite number of variational
parameters. As a result, the thermodynamic functions obtained by the TPVA
show mean-field-like behavior around $T_{\rm c}$. Thus, in order to
obtain critical indices from the numerical data by the TPVA, one has
to calculate the thermodynamic functions in off critical region, where
the TPVA is more accurate. 

The main purpose of this paper is to obtain the magnetic critical exponent
$\beta$ for the 3D lattice models using the spontaneous magnetization calculated
in the off critical region. We assume a scaling form for the numerical fitting
of $\beta$ with respected to the calculated data.

The paper is organized as follows: in Section~2, we define the model
and discuss the variational background of the numerical method. We
propose four types of tensor product states (TPS) which are used to
calculate the variational state. By increasing the total number of
variational parameters in the TPS, we obtain more precise numerical
results. We calculate the critical temperature and the magnetic
critical exponent for these four types of TPS in Section~3.
A brief summary of the results obtained is summarized in Section~4.

\section{Model and variational background}

We consider the 3D Ising model on a simple cubic lattice of infinite size
along all the $x$, $y$, and $z$ axis coordinates. The model studied is
described by the spin Hamiltonian
\begin{equation}
{\cal H} = -J\sum\limits_{i,j,k}\sigma_{i,j,k}\left(
\sigma_{i+1,j,k}\,+\,\sigma_{i,j+1,k}\,+\,\sigma_{i,j,k+1}\right)
\end{equation}
with the nearest-neighbor spins $\sigma=\pm1$ interacting via coupling
constant $J$. The transfer matrix ${\cal T}$ connects
two adjacent spin layers $[\sigma_k]$ and $[\sigma_{k+1}]$. It can
be expressed by product of the Boltzmann weights
\begin{equation}
{\cal T}[\sigma_k|\sigma_{k+1}]=\prod\limits_{i,j}
W^{\rm B}_{i,j}\{\sigma_k|\sigma_{k+1}\}\ ,
\end{equation}
where we abbreviated the notation gathering four spins into one
\begin{equation}
\{\sigma_k\}\equiv(\sigma_{i,j,k}\ \sigma_{i+1,j,k}
\ \sigma_{i,j+1,k}\ \sigma_{i+1,j+1,k})\ .
\end{equation}
We have employed the local Boltzmann weight of the interaction round-a-face
(IRF) type which is defined as
\begin{eqnarray}
\nonumber
W^{\rm B}_{i,j}\{\sigma_k|\sigma_{k+1}\}=\exp\left[\frac{J}{4k_{\rm B}T}(
 \sigma_{i,j,  k}\sigma_{i,j,  k+1}+\sigma_{i+1,j,  k}\sigma_{i+1,j,  k+1}
\right.
\end{eqnarray}
\vspace{-0.7cm}
\begin{eqnarray}
\nonumber
 \phantom{\frac{1}{1}}
 + \sigma_{i,j+1,k}\sigma_{i,j+1,k+1}+\sigma_{i+1,j+1,k}\sigma_{i+1,j+1,k+1}
 + \sigma_{i,  j,  k}\sigma_{i+1,j,  k}+\sigma_{i+1,j,  k}\sigma_{i+1,j+1,k}
\end{eqnarray}
\vspace{-0.7cm}
\begin{eqnarray}
\nonumber
 \phantom{\frac{1}{1}}
 + \sigma_{i+1,j+1,k}\sigma_{i,  j+1,k}+\sigma_{i,  j+1,k}\sigma_{i,  j,  k}
 + \sigma_{i,  j,  k+1}\sigma_{i+1,j,  k+1}+\sigma_{i+1,j,  k+1}\sigma_{i+1,j+1,k+1}
\end{eqnarray}
\vspace{-0.7cm}
\begin{eqnarray}
 \left.\phantom{\frac{1}{1}}
+ \sigma_{i+1,j+1,k+1}\sigma_{i,  j+1,k+1}+\sigma_{i,  j+1,k+1}\sigma_{i,  j,  k+1})
\right]\ ,
\end{eqnarray}
where $k_{\rm B}$ is the Boltzmann constant and $T$ is temperature.
Due to simplicity, we set $J/k_{\rm B}=1$ in what follows on.

The variational partition function per layer for the given transfer
matrix ${\cal T}$ has the form
\begin{equation}
\lambda_{\rm var}(\Psi)=\frac{\langle\Psi|{\cal T}|\Psi\rangle}
{\langle\Psi|\Psi\rangle}=\frac{\sum\limits_{[\sigma_k],[\sigma_{k+1}]}\Psi[\sigma_k]
{\cal T}[\sigma_k|\sigma_{k+1}]\Psi[\sigma_{k+1}]}
{\sum\limits_{[\sigma_k]}(\Psi[\sigma_k])^2}.
\label{vpf}
\end{equation}
Purpose of the TPVA is to maximize Eq.~(\ref{vpf}) by a proper approximation
of the trial state $\Psi$. We assume that $\Psi$ can be written
by the product of identical local variational weights $V$. This approximation
is often referred to as the tensor product state~\cite{TPVA4,Nishio,Ver1,Ver2,Ver3}.

In the following, we consider four candidates of $\Psi$, where the difference
is in the definition of the local weights $V$ and their connection.
\begin{figure}[tbp]
\begin{center}
\includegraphics[width=10cm,clip]{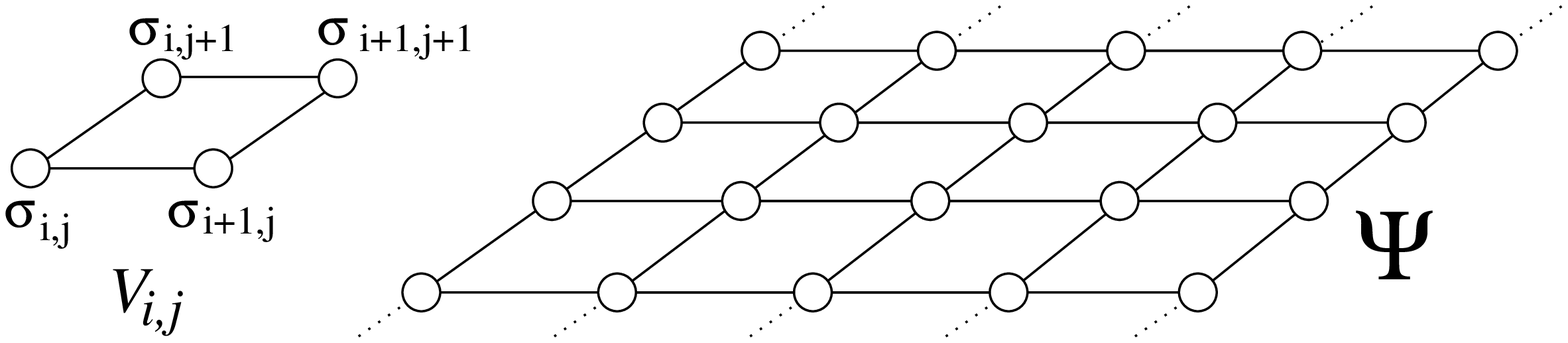}
\end{center}
\caption{Graphical representation of the four-spin local variational weight
$V_{i,j}$ (left) which is used to construct the trial state $\Psi$ (right).
Notice that we have dropped the subscript $k$ from the spins for simplicity.}
\label{fig1}
\end{figure}
\begin{figure}[tbp]
\begin{center}
\includegraphics[width=10cm,clip]{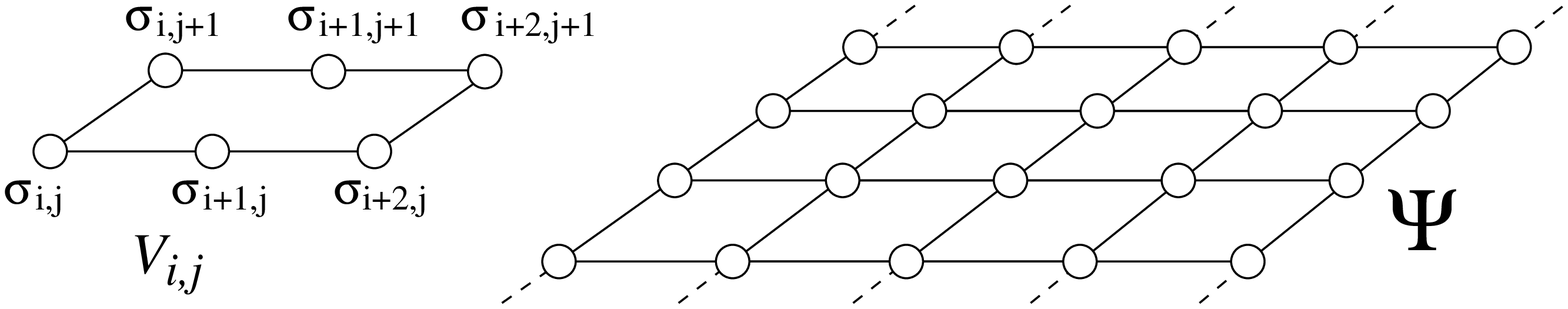}
\end{center}
\caption{The six-spin local variational weight $V_{i,j}$ (left) and
the trial state $\Psi$ (right). The weights $V_{i,j}$ overlap one another
only in the $x$ axis direction (denoted by the subscript $i$).}
\label{fig2}
\end{figure}
\begin{itemize}
\item[(i)] In the lowest approximation, the single local weight $V$ consists
of four spins. Hence, we have $2^4=16$ adjustable parameters and $V$ graphically
represents a square-shaped object as shown in Fig.~\ref{fig1}. Taking product
over all the uniformly given local weights $V$, the trial state yields
\begin{equation}
\Psi[\sigma_k]
=\prod\limits_{i,j}V_{i,j}(\sigma_{i,j,k},\sigma_{i+1,j,k},\sigma_{i,j+1,k}
\sigma_{i+1,j+1,k}).
\end{equation}
\item[(ii)] A way to increase precision of the variational calculations is to 
enlarge the size of the local weight $V$ up to six spins, see Fig.~\ref{fig2}.
Then, we let the weights $V$ partially overlap one another defining the trial
state as follows
\begin{equation}
\Psi[\sigma_k]
=\prod\limits_{i,j}V_{i,j}(\sigma_{i,j,k},\sigma_{i+1,j,k},\sigma_{i+2,j,k}
\sigma_{i,j+1,k},\sigma_{i+1,j+1,k},\sigma_{i+2,j+1,k}).
\end{equation}
The number of the adjustable parameters is $2^6=64$.
\begin{figure}[htbp]
\begin{center}
\includegraphics[width=10cm,clip]{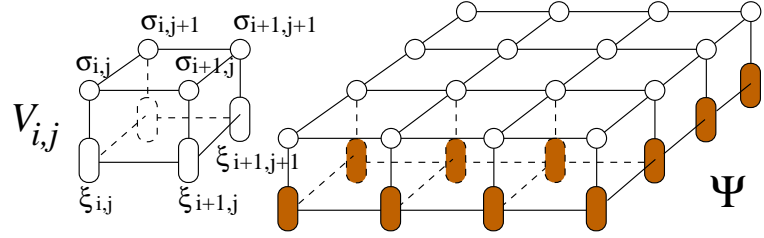}
\end{center}
\caption{The local variational weight $V_{i,j}$ of the IRF type with the
auxiliary variables $\xi$ (left) and the trial state $\Psi$ (right).
The dark ovals represents the auxiliary variables to be summed up.}
\label{fig3}
\end{figure}
\begin{figure}[htbp]
\begin{center}
\includegraphics[width=10cm,clip]{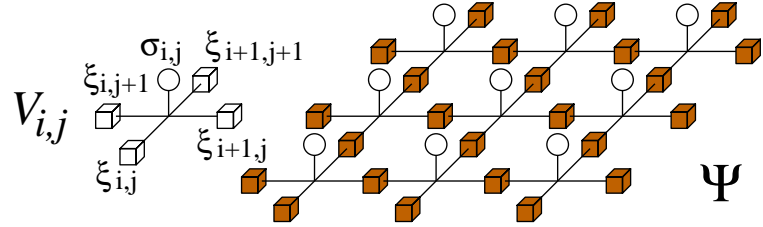}
\end{center}
\caption{The local variational weight $V_{i,j}$ of the vertex type with the
auxiliary variables $\xi$ (left) and the trial state $\Psi$ (right).
The dark cubes are the summed up auxiliary variables.}
\label{fig4}
\end{figure}
\item[(iii)] Further increase of the free parameters leads to a better approximation
of the trial state (for details, see Ref.~\cite{TPVA4}). Therefore, we introduce
the so-called auxiliary variables which are included in the local weight $V$
\begin{equation}
\Psi\left[ \sigma_k \right ]
\, = \, \sum\limits_{[\xi]}\prod\limits_{i,j}
V\left( {{\sigma_{i,j,k}^{~}}\atop{\xi_{i,j,k}^{~}}}
\ {{\sigma_{i+1,j,  k}}\atop{\xi_{i+1,j,  k}}}
\ {{\sigma_{i,  j+1,k}}\atop{\xi_{i,  j+1,k}}}
\ {{\sigma_{i+1,j+1,k}}\atop{\xi_{i+1,j+1,k}}} \right),
\label{vsIRF}
\end{equation}
where $\xi$ denotes the auxiliary variable which can be in one of
$n$ states $(1,2,...,n)$. The sum runs over all the auxiliary
variables as depicted in Fig.~\ref{fig3}. Thus, there are $(2n)^4$
adjustable parameters in total. By setting $n=1$, the case (i) is satisfied.
\item[(iv)] Finally, we consider the vertex-type representation~\cite{CTMRG1}
of the 3D Ising model where the auxiliary variables $\xi$ can be naturally
included. The trial state is then given by
\begin{equation}
\Psi\left[ \sigma_k \right ]
\, = \, \sum\limits_{[\xi]}\prod\limits_{i,j}
V\left( {\sigma_{i,j,k}}\atop{\xi_{i,j,k}
\ \ \xi_{i+1,j,  k}\ \ \xi_{i,  j+1,k}\ \ \xi_{i+1,j+1,k}} \right)
\label{vsvertex}
\end{equation}
and is graphically represented in Fig.~\ref{fig4}. There are
$2n^4$ parameters in total.
\end{itemize}

We treat the local variational weights $V$ with different number
of adjustable parameters (up to 256) by the TPVA.
In fact, not all of them are independent due to the symmetries
of the model. The main advantage of the numerical algorithm
TPVA is that we have derived a self-consistent equation (based
on the a generalized eigenvalue problem) which iteratively
tunes all the adjustable parameters regardless of initial
choice. Numerical details of the self-consistent improvement
for the trial state $\Psi$ with explanation of the TPVA are
reviewed in Refs.~\cite{TPVA1,TPVA2,TPVA3,TPVA4,Nishio,ANNNI}.

Having written the trial state $\Psi$ and the transfer matrix
${\cal T}$ in the product forms, we can accurately calculate
the numerator of Eq.~(\ref{vpf})
\begin{equation}
\langle\Psi|{\cal T}|\Psi\rangle=
\sum\limits_{[\sigma_k],[\sigma_{k+1}]}\ \ \prod\limits_{i,j}
V_{i,j}\{\sigma_k\}
W^{\rm B}_{i,j}\{\sigma_k|\sigma_{k+1}\}V_{i,j}\{\sigma_{k+1}\}
\label{eq1}
\end{equation}
and also the denominator
\begin{equation}
\langle\Psi|\Psi\rangle=
{\sum\limits_{[\sigma_k]}\ \ \prod\limits_{i,j}
\left(V_{i,j}\{\sigma_k\}\right)^2}
\label{eq2}
\end{equation}
using the renormalization techniques.
In particular, we use both the DMRG and its variant the
Corner Transfer Matrix Renormalization Group~\cite{Nishino,CTMRG1,CTMRG2}.

\section{Results}

\begin{figure}[tb]
\begin{center}
\includegraphics[width=6.6cm,clip]{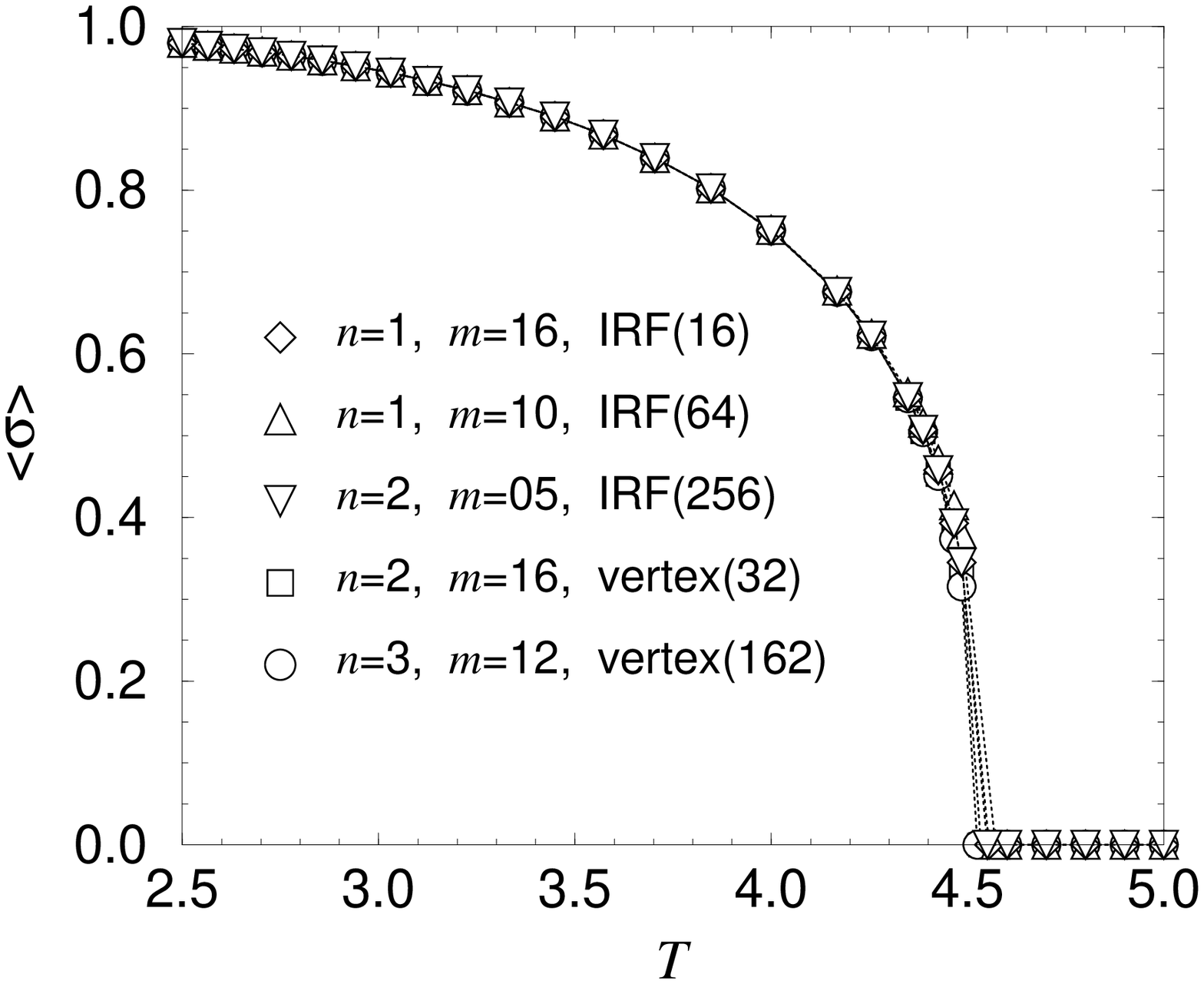}
\includegraphics[width=6.6cm,clip]{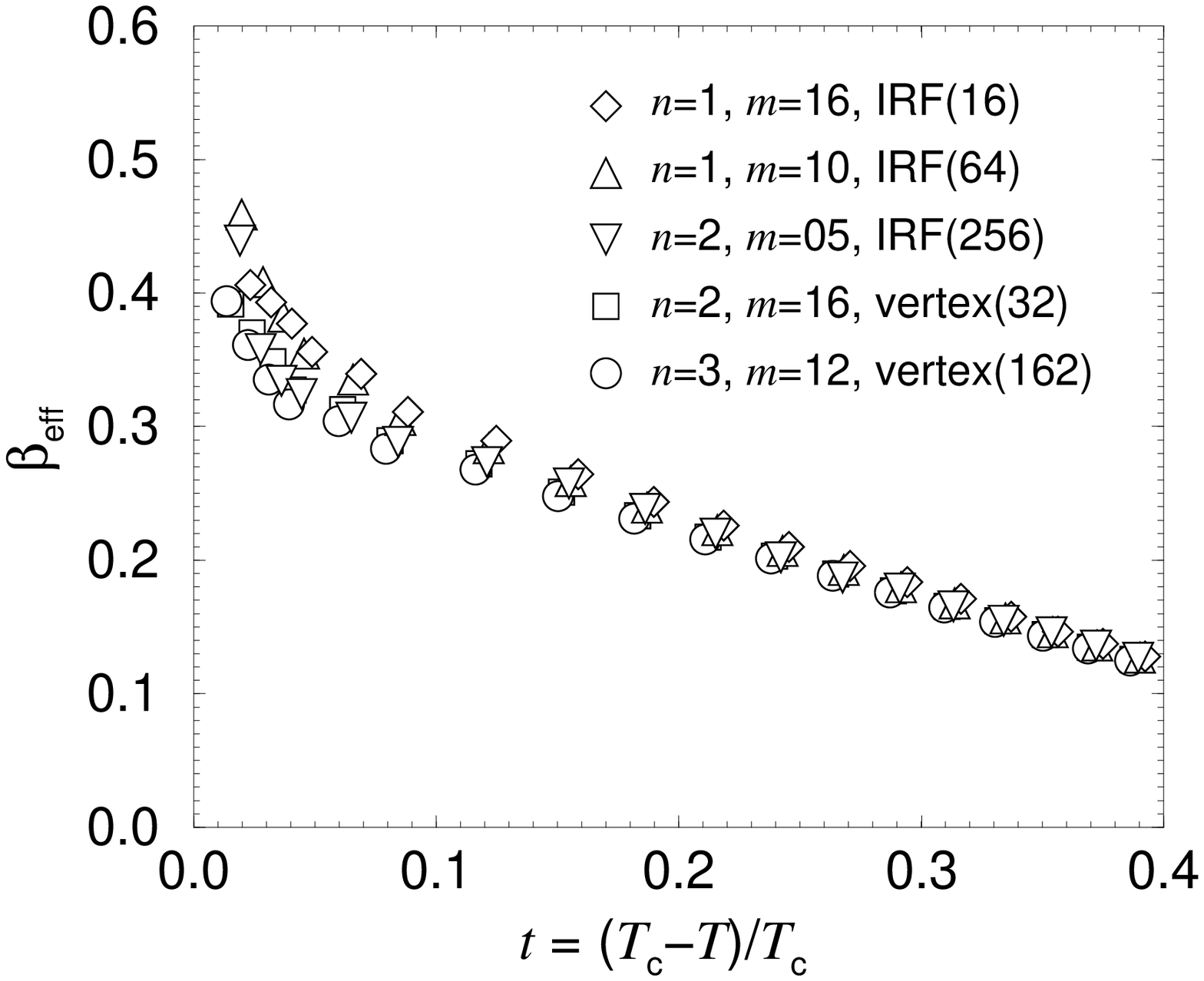}
\end{center}
\caption{Left: dependence of the spontaneous magnetization $\langle\sigma\rangle$
on temperature $T$ for selected variables $n$ and $m$ as listed in Table~1.
Right: the plot of the effective critical exponent $\beta_{\rm eff}$ versus relative
temperature $t$ for the same variables as shown on the left graph. The total
number of the adjustable parameters is shown in the parenthesis.}
\label{fig56}
\end{figure}

We calculate the spontaneous magnetization per spin site using the relation
\begin{equation}
\langle\sigma\rangle\equiv\langle\sigma_{i,j}\rangle=\frac{\langle\Psi|\sigma_{i,j}|\Psi\rangle}
{\langle\Psi|\Psi\rangle}
\label{mag}
\end{equation}
for all the four kinds of approximation discussed above. The corresponding
graph is shown in Fig.~\ref{fig56} on the left. The number of the
states kept in the DMRG's block-spin variable is denoted by the letter
$m$. Details which concern the meaning of this $m$-state variable
can be found in Ref.~\cite{White1,White2,DMRG,Nishino}.
The Table~1 summarizes the critical temperatures $T_{\rm c}$ obtained
by the TPVA for the several selected variables $n$ and $m$.
The MC result ($T_{\rm MC}=4.5116$) is used as
a reference to evaluate the corresponding relative error
${\cal E}_{\rm T}=(T_{\rm c}-T_{\rm MC})/T_{\rm MC}$.
Precision of our results improves with the
increasing number of the adjustable parameters in $V$ (for both the
IRF and the vertex type lattice models).

We calculate the magnetic critical exponent $\beta$ which describes the
vanishing of the spontaneous magnetization
\begin{equation}
\label{betaexp}
\langle\sigma\rangle \propto {\left(\frac{T_{\rm c}-T}{T_{\rm c}}\right)}^\beta
\equiv t^\beta
\end{equation}
when approaching the critical temperature $T_{\rm c}$ from the ferromagnetic phase.
We analyze the effective critical exponent $\beta_{\rm eff}$ as a function
of the relative temperature $t$~\cite{betaexp1,betaexp2}
\begin{equation}
\beta_{\rm eff}(t)=\frac{\rm d}{{\rm d}\log(t)}\log\langle\sigma(t)\rangle.
\end{equation}

\begin{table}[t]
\begin{center}
\begin{tabular}{|c|r|c|c|c|c|c|c|}
\hline
$n$ & $m$ & type of TPS & lattice type &  $T_{\rm c}$ & ${\cal E}_{\rm T}$ / [\%]
& $\beta$ & ${\cal E}_{\beta}$ / [\%]\\
\hline
\hline
1 & 16 & (i) &  IRF   & 4.570 & 1.30 & 0.357 & 9.58 \\
\hline
1 & 10 & (ii)&  IRF   & 4.554 & 0.94 & 0.348 & 6.81 \\
\hline
2 &  5 &(iii)&  IRF   & 4.550 & 0.85 & 0.338 & 3.74 \\
\hline
2 & 16 & (iv)& vertex & 4.533 & 0.47 & 0.332 & 1.90 \\
\hline
3 & 12 & (iv)& vertex & 4.525 & 0.30 & 0.327 & 0.37 \\
\hline
\end{tabular}
\end{center}
\medskip
\caption {List of the critical temperatures $T_{\rm c}$ and the magnetic
critical exponents $\beta$ computed for the 3D Ising model for various
kinds of approximation. The relative errors ${\cal E}_{\rm T}$ and
${\cal E}_{\beta}$ are computed with respect to the results of the MC
simulations~\cite{MC}.}
\end{table}

Figure~\ref{fig56} (on the right) displays dependence of the effective
exponent $\beta_{\rm eff}$ on the reduced temperature $t$. It is evident
that for $t<0.1$ the linear dependence of the exponent $\beta_{\rm eff}$
is broken and $\beta_{\rm eff}$ tends to the mean-field
value equal to 1/2.

\begin{figure}[tb]
\begin{center}
\includegraphics[width=9cm,clip]{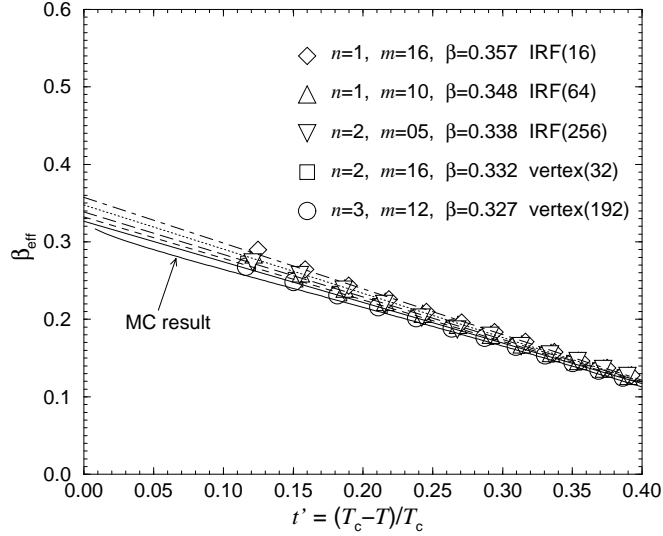}
\end{center}
\caption{The extrapolation of the effective critical exponent
$\beta_{\rm eff}(t^\prime)$ down to $t^\prime=0$ using Eq.~({\ref{fitting}}).
The plotted numerical data are not affected by the mean-field effect.
The lowest full line is obtained from the fit of the MC simulation~\cite{MCbeta}.}
\label{FIG7}
\end{figure}

In order to calculate $\beta$, we assume dependence of the spontaneous
magnetization $\langle\sigma\rangle$ on the relative temperature $t$
in the series
\begin{equation}
\langle\sigma(t)\rangle=\gamma t^{\beta_{\rm eff}(t)}
=\gamma t^{(\beta+\alpha_1 t+\alpha_2 t^2+\alpha_3 t^3+\cdots)}
\end{equation}
with unknown constants $\gamma$, $\beta$, and $\alpha_i$ for $i=1,2,\dots$.
Neglecting the terms of the second-order and higher in the effective
exponent $\beta_{\rm eff}$, we have a linear dependence. The linear
dependence is well satisfied in the case of the 2D Ising model~\cite{betaexp2}.
For simplicity, we also assume the same linear dependence for the 3D Ising model~\cite{betaexp1},
in particular,
\begin{equation}
\langle\sigma(t^\prime)\rangle=\gamma t^{\prime\ \beta_{\rm eff}(t^\prime)}
=\gamma t^{\prime\ \beta+\alpha_1 t^\prime}
\label{fitting}
\end{equation}
with  $t^\prime=(T_{\rm c}-T)/T_{\rm c}$ where $T_{\rm c}$ is obtained
by the TPVA as listed in Table~1. Note, that $\beta_{\rm eff}(t^\prime=0)=1/2$
is the consequence of the mean-field behavior. Therefore, we use
the linear extrapolation of the effective critical exponent
$\beta_{\rm eff}(t^\prime)=\beta+\alpha_1 t^\prime$ for $t^\prime\gg0$
within the off-critical region in order to find the constants $\beta$
and $\alpha_1$. We identify the constant $\beta$ with the critical
exponent after taking the limit $t^\prime\to0$ from the linear fit.
The results are depicted in Fig.~\ref{FIG7}. Justification to carry
out the linear extrapolation is supported by the plotting of the MC
result~\cite{MCbeta} in the same graph which gives the nearly linear
behavior. The critical exponents $\beta$ obtained for various variables
$n$ and $m$ are listed in Table~1 and compared with the MC result
$\beta_{\rm MC}=0.3258(14)$~\cite{MCexp}.

The error in determining $\beta$ from the extrapolation procedure is
deduced from the following consideration. Assume a small deviation
$\varepsilon$ from the critical temperature $T_{\rm c}$ obtained by the
TPVA, i.~.e, $T_{\rm c}^\prime=T_{\rm c}+\varepsilon$. Table~2 shows that
for a given $\varepsilon$, the relative error in $T_{\rm c}$ affects the
relative error in $\beta$ of about ten times.

\begin{table}[!ht]
\begin{center}
\begin{tabular}{|r|c|c|c|c|}
\hline
$\varepsilon$\ \ \ \ \  & $T_{\rm c}^\prime(\varepsilon)$ & ${\cal E}_{\rm T^\prime}$ / [\%]
& $\beta$ & ${\cal E}_{\beta(\varepsilon)}$ / [\%]\\
\hline
\hline
-0.010 & 4.515  &  0.22   &   0.320  &   2.1   \\
\hline
-0.005 & 4.520  &  0.11   &   0.323  &   1.1   \\
\hline
 0.000 & 4.525  &  ---    &   0.327  &   ---   \\
\hline
+0.005 & 4.530  &  0.11   &   0.330  &   1.1   \\
\hline
+0.010 & 4.535  &  0.22   &   0.334  &   2.2   \\
\hline
\end{tabular}
\end{center}
\medskip
\caption {The relative errors ${\cal E}_{\rm T^\prime}$ and
${\cal E}_{\beta(\varepsilon)}$ calculated for given deviations $\varepsilon$
from the $T_{\rm c}$. This example is demonstrated on the data with
the variables $n=3$ and $m=12$.}
\end{table}

\section{Conclusion}

We have applied the TPVA to the simple 3D Ising model and calculated
the critical temperature from the spontaneous magnetization. Using the
scaling form, we obtained the magnetic critical exponent $\beta$ from
the spontaneous magnetization in the off critical region. The way of
estimation of the critical index is of use for future applications of
the TPVA to various 3D classical lattice models.

\begin{ack}
A.G. is supported by the VEGA grant No. 2/3118/23.
\end{ack}

\end{document}